\documentclass[12pt]{article}

\usepackage{latexsym,amsmath,amssymb,theorem,epsfig}

\DeclareFontFamily{OT1}{rsfs10}{}
\DeclareFontShape{OT1}{rsfs10}{m}{n}{ <-> rsfs10 }{}
\DeclareMathAlphabet{\mathscript}{OT1}{rsfs10}{m}{n}

\numberwithin{equation}{section}

\newcommand{\ns}{\normalsize}

\theoremstyle{plain}

{\theorembodyfont{\rmfamily} }

\begin{document}

%%%%%%%%%%%%%%%%%%%%%%%%%%%%%%%%%%%%%%%%%%%%%%%%%%%%%%%%%%%%%%%%%%%%%%

\begin{titlepage}

\vspace{-5cm}

\title{
  \hfill{\ns }  \\[1em]
   {\LARGE Derivative F-Terms from Heterotic M-Theory Five-brane Instantons}
\\[1em] }
\author{
   Evgeny I. Buchbinder
     \\[0.5em]
   {\ns Perimeter Institute for Theoretical Physics} \\[-0.4cm]
{\ns Waterloo, Ontario, N2L 2Y5, Canada}\\[0.3cm]}

\date{}

\maketitle

\begin{abstract}

We study non-perturbative effects due to a 
heterotic M-theory five-brane wrapped on Calabi-Yau 
threefold. 
We show that such instantons contribute 
to derivative F-terms described recently by Beasley and Witten 
rather than to the superpotential.

\end{abstract}

\thispagestyle{empty}

\end{titlepage}

%%%%%%%%%%%%%%%%%%%%%%%%%%%%%%%%%%%%%%%%%%%%%%%%%%%%%%%%%%%%%%%%%%%%%%%%%%%%%%%%%%%%%%%%%%%%%%%%%%%%%%%%

\section{Introduction}

%%%%%%%%%%%%%%%%%%%%%%%%%%%%%%%%%%%%%%%%%%%%%%%%%%%%%%%%%%%%%%%%%%%%%%%%%%%%%%%%%%%%%%%%%%%%%%%%%%%%%%

In~\cite{Beasley1} and~\cite{Beasley2} Beasley and Witten described 
a new class of instanton effects in ${\cal N}=1$ supersymmetric 
field theory and string theory. 
These effects are generating of derivative, or multi-fermion, F-terms. 
In superfields, the corresponding contributions to the effective action are  
$F$-terms containing spinor covariant derivatives.
In components, this corresponds to four- and higher-fermionic terms. 
The simplest non-trivial example of such terms is two-derivative, or 
four-fermion, F-terms. They correspond to
non-perturbative deformation of the classical
moduli space. A field theory example where 
such effects take place is ${\cal N}=1$ supersymmetric QCD 
with $N_f=N_c$~\cite{Seiberg1, Seiberg2}. 
In general, as was shown in~\cite{Beasley1}, for $N_f \geq N_c$ 
instantons contribute to interactions of $2 (N_f-N_c)+4$ fermions. 
In the case of string theory, 
it was shown in~\cite{Beasley2} that derivative F-terms arise, in particular, 
from wrapping heterotic strings on non-isolated
genus zero or higher genus curves. It is worth pointing out that, 
just like superpotential, derivative F-terms have a certain holomorphic signature.

In this note, we present an explicit example of a string instanton
that contributes to four-fermion terms in the low-energy effective action.
The instanton is a heterotic M-theory five-brane wrapped on
Calabi-Yau threefold. Being a bulk instanton, a five-brane wrapping 
Calabi-Yau threefold has four fermionic zero modes.
As the result
it cannot contribute 
to the superpotential.  
Instead, we show that it contributes to derivative F-terms.
This situation is different from the case of string or open membrane 
instantons~\cite{Becker, Wittenpoten, Wittend, Lima1, Lima2, Donagi1, Donagi2}. These instantons
are located on the boundary of the interval and, thus, have less 
fermionic zero modes. In fact, they have only two zero modes and contribute
to the superpotential for various moduli. 

This note is organized as follows. In section 2, we give a brief review 
of the structure of derivative F-terms. In section 3, we study 
the action of a bulk five-brane instanton wrapping Calabi-Yau manifold.
We show that this instanton contributes to four-fermion F-terms.
In section 4, we discuss some additional features of five-brane instantons. 

%%%%%%%%%%%%%%%%%%%%%%%%%%%%%%%%%%%%%%%%%%%%%%%%%%%%%%%%%%%%%%%%%%%%%%%%%%%%%%%%%%%%%%%%%%%%%%%%%%%%%%%%%

\section{Review of Derivative F-terms}

%%%%%%%%%%%%%%%%%%%%%%%%%%%%%%%%%%%%%%%%%%%%%%%%%%%%%%%%%%%%%%%%%%%%%%%%%%%%%%%%%%%%%%%%%%%%%%%%%%%%%%%

In this section, we give a brief review of the derivative F-terms 
following~\cite{Beasley1, Beasley2}. 
These terms have the following superfield structure 
\begin{equation}
\delta S=\int d^4 x {\ }d^4 \theta {\ }\omega_{{\bar i}{\bar j}}(\Phi, \bar \Phi)
{\bar D}_{\dot \alpha}{\bar \Phi}^{{\bar i}}
{\bar D}^{\dot \alpha}{\bar \Phi}^{{\bar j}},
\label{1}
\end{equation}
where $\Phi^i$ and ${\bar \Phi}^{{\bar i}}$ are
chiral and anti-chiral superfields whose lowest components 
$\phi^i$ and ${\bar \phi}^{{\bar i}}$ parametrize the classical moduli space ${\cal M}_{cl}$.
Furthermore, $\omega_{{\bar i}{\bar j}}$ is defined as
\begin{equation}
\omega_{{\bar i}{\bar j}}=\frac{1}{2}\left(g_{i {\bar i}}{\ }\omega_{{\bar j}}^{{\ }i}+
g_{i {\bar j}}{\ }\omega_{{\bar i}}^{{\ }i}\right),
\label{2}
\end{equation}
where $g_{i \bar i}$ is the Kahler metric on ${\cal M}_{cl}$ and $\omega_{{\bar j}}^{{\ }i}$
represents an element in the Dolbeault cohomology group 
$H^1_{\bar \partial}({\cal M}_{cl}, T{\cal M}_{cl})$. Thus, $\omega_{{\bar j}}^{{\ }i}$ 
is independent of the metric and parametrizes infinitesimal deformations of the complex 
structure of ${\cal M}_{cl}$. Let us note that it is possible
to choose a coordinate system on ${\cal M}_{cl}$ in which the 
antiholomorphic components of the Christoffel symbols, $\Gamma_{{\bar i}{\bar j}}^{\bar k}$ vanish. 
In this coordinate system, $\omega_{{\bar i}{\bar j}}$ becomes holomorphic, that is
a function of $\Phi^i$ only. 
The operator $\omega_{{\bar i}{\bar j}}
{\bar D}_{\dot \alpha}{\bar \Phi}^{{\bar i}}
{\bar D}^{\dot \alpha}{\bar \Phi}^{{\bar j}}$ 
is not manifestly chiral. Its chirality follows from the equations of motion 
of the unperturbed sigma-model with the metric $g_{i \bar i}$ and holomorphy 
of $\omega_{{\bar j}}^{{\ }i}$. 
Being an element of $H^1_{\bar \partial}({\cal M}_{cl}, T{\cal M}_{cl})$
rather than a function, $\omega_{{\bar j}}^{{\ }i}$ 
is locally exact. Therefore, locally, one can always rewrite $\delta S$ as 
an integral over the whole superspace. However, if the cohomology class
represented by $\omega_{{\bar j}}^{{\ }i}$ is non-trivial, we cannot 
write $\delta S$ globally as a D-term. In this sense, $\delta S$ 
represents an F-term. 
The expression in~\eqref{1} can be generalized to higher derivative F-terms 
\begin{equation}
\delta S=\int d^4 x {\ }d^4 \theta {\ }\omega_{{\bar i}_1\dots {\bar i}_p
{\bar j}_1 \dots {\bar j}_p}
\left({\bar D}_{\dot \alpha}{\bar \Phi}^{{\bar i}_1}
{\bar D}^{\dot \alpha}{\bar \Phi}^{{\bar j}_1}\right) \dots
\left({\bar D}_{\dot \alpha}{\bar \Phi}^{{\bar i}_p}
{\bar D}^{\dot \alpha}{\bar \Phi}^{{\bar j}_p}\right).
\label{3}
\end{equation}
See~\cite{Beasley1, Beasley2} for details.

In components, the action~\eqref{1} becomes
\begin{equation}
\delta S= -4\int d^4x {\ } \omega_{{\bar i}{\bar j}} {\ }\partial_{\mu}{\bar \phi}^{{\bar i}}
\partial^{\mu}{\bar \phi}^{{\bar j}} -\frac{1}{4}
\int d^4 x {\ }\triangledown_k \triangledown_l {\ }\omega_{{\bar i}{\bar j}}{\ } 
\psi^{k \alpha}\psi^l_{\alpha} {\bar \psi}^{\bar i}_{\dot \alpha} {\bar \psi}^{{\bar j} \dot \alpha}
+\dots {\ },
\label{4}
\end{equation}
where $\psi^i_{\alpha}$ is the fermionic superpartner of $\phi^i$. 
%To obtain the second 
%term in~\eqref{3} one needs to shift the auxiliary field $F^i$ by 
%$-\Gamma_{jk}^i \psi^{j \alpha}\psi^{k}_{\alpha}$ as in the 
%case of supersymmetric sigma-model. 

The most convenient way to see if an instanton contributes to $\delta S$
is to look at the correlation function of four fermions
\begin{equation}
<\psi^{k \alpha}(x_1)
\psi^l_{\alpha}(x_2) 
{\bar \psi}^{\bar i}_{\dot \alpha}(x_3) {\bar \psi}^{{\bar j} \dot \alpha}(x_4)>.
\label{5}
\end{equation}
In order for this correlation function to be non-zero, the action of the 
instanton has to be multiplied by four fermions. This implies that the instanton 
has to have four fermionic zero modes. In the next section, we will present
an example of such an instanton, a heterotic M-theory five-brane wrapped 
on a Calabi-Yau manifold. The general expression for the correlation function that 
we want to consider is
\begin{eqnarray}
&&<\psi^{k \alpha}(x_1)
\psi^l_{\alpha}(x_2) 
{\bar \psi}^{\bar i}_{\dot \alpha}(x_3) {\bar \psi}^{{\bar j} \dot \alpha}(x_4)>=\nonumber \\
&&\int{\cal D}{\Phi}{\bar {\cal D}}{\bar \Phi}
e^{-S_{4D}}
\psi^{k \alpha}(x_1)
\psi^l_{\alpha}(x_2) 
{\bar \psi}^{\bar i}_{\dot \alpha}(x_3) {\bar \psi}^{{\bar j} \dot \alpha}(x_4)
\int {\cal D}{\mathbb Z}e^{-S_5},
\label{5.5}
\end{eqnarray}
where by $S_{4D}$ we denote the four-dimensional action
and $\int {\cal D}{\mathbb Z}e^{-S_5}$ is the partition function 
of the five-brane wrapped on Calabi-Yau manifold.

%%%%%%%%%%%%%%%%%%%%%%%%%%%%%%%%%%%%%%%%%%%%%%%%%%%%%%%%%%%%%%%%%%%%%%%%%%%%%%%%%%%%%%%%%%%%%%%%%%%%%%%

\section{Five-brane Instantons}

%%%%%%%%%%%%%%%%%%%%%%%%%%%%%%%%%%%%%%%%%%%%%%%%%%%%%%%%%%%%%%%%%%%%%%%%%%%%%%%%%%%%%%%%%%%%%%%%%%

We consider a (Euclidean) five-brane wrapped on a Calabi-Yau manifold and located at an 
arbitrary point in the interval away from any of the orbifold fixed planes. 
The relevant part of the five-brane action~\cite{Sorokin} is 
\begin{equation}
S_5 =T\int d^6 \sigma \sqrt{detg_{ij}} + iT\int {\mathbb C}^{(6)}.
\label{6}
\end{equation}
Here $\sigma^i, i=1, \dots, 6$ are the coordinates along the five-brane
worldvolume, $T$ is the five-brane tension, $g_{ij}$ is the pullback of the 
metric superfield to the worldvolume and ${\mathbb C}^{(6)}$ is the superfield whose
lowest component is the dual six-form potential $C_{M_1 \dots M_6}$. 
More precisely. 
$g_{ij}$ is defined as 
\begin{equation}
g_{ij}=\eta_{AB}{\mathbb E}_{{\mathbb M}}^A {\mathbb E}_{{\mathbb N}}^B 
\partial_i {\mathbb Z}^{{\mathbb M}} \partial_j {\mathbb Z}^{{\mathbb N}}
=g_{{\mathbb M} {\mathbb N}}\partial_i {\mathbb Z}^{{\mathbb M}} 
\partial_j {\mathbb Z}^{{\mathbb N}}.
\label{7}
\end{equation}
Our index convention is the following. The indices $A, B$ are flat and run from $0$ to 
$10$. The superindex ${\mathbb M}$ is split into the space-time index $M, M=0, \dots 10$
and the spinor index $m, m=1, \dots 32$. The index $M$ itself splits into the 
four-dimensional index $\mu, \mu=0, 1, 2 ,3$, the indices along the Calabi-Yau threefold 
$U, U=4, \dots, 10$ and the eleventh direction. 
The supercoordinates ${\mathbb Z}^{{\mathbb M}}$ split into the space-time 
coordinates $X^{M}$ and fermionic coordinates $\Theta^{m}$.
To proceed we have to expand the superfields ${\mathbb E}_{{\mathbb M}}^A$ and
${\mathbb C}^{(6)}$ in powers in the fermionic coordinates $\Theta$. For our purposes
it is enough to expand to the linear order in the gravitino $\Psi_M$
similarly to the case of membrane instantons in~\cite{Lima1, Lima2}.
We have~\cite{deWit}
\begin{eqnarray}
&&{\mathbb E}_{M}^A=E_M^A+2{\bar \Theta}\Gamma^A \Psi_M+\dots{\ }, \nonumber \\
&&{\mathbb E}_{m}^A=0, \nonumber \\
&&{\mathbb C}_{M_1M_2M_3M_4M_5M_6}^{(6)}= C_{M_1M_2M_3M_4M_5M_6}-
{\bar \Theta}\Gamma_{[M_1M_2M_3M_4M_5}\Psi_{M6]}+\dots {\ }.
\label{8}
\end{eqnarray}
%
%Let us point out that in this paper, for simplicity, we will consider only 
%the instanton action evaluated on the solution of the equations of motion
%and will not consider the one-loop determinants that arise after expanding the 
%action around the solution. To find the determinants one has to expand 
%in~\eqref{8} to the second order in $\Theta$. The calculations will be analogous
%to the ones in~\cite{Lima1, Lima2}. We will not do them in this paper. 
%However, some of the one-loop determinants can be obtained 
Substituting eq.~\eqref{8} back in the action and expanding to the linear 
order in gravitino yields
\begin{eqnarray}
S_5 &=&  T\int d^6 \sigma 
\sqrt{detg_{MN}\partial_i X^M \partial_j X^N}+\frac{iT}{6!} \int d^6 \sigma
\epsilon^{i_1 \dots i_6}\partial_{i_1}X^{M_1}\dots 
\partial_{i_6}X^{M_6}C_{M_1 \dots M_6} \nonumber \\
    &+& T\int d^6 \sigma 
\sqrt{detg_{MN}\partial_i X^M \partial_j X^N}V^N\Psi_N,
\label{9}
\end{eqnarray}
where the vertex operator for the gravitino is given by
\begin{equation}
V^N=g^{ij}\partial_i X^M \partial_j X^N (\bar \Theta \Gamma_M)-
\frac{\epsilon^{i_1 \dots i_5 i}}{6!\sqrt{g_{ij}}}
\partial_{i_1}X^{M_1}\dots \partial_{i_5}X^{M_5} \partial_{i}X^{N}
(\bar \Theta \Gamma_{{M_1}\dots M_5}).
\label{10}
\end{equation}

Now let us recall that we are considering a five-brane wrapped on a 
Calabi-Yau threefold. From the four-dimensional point of view this 
configuration looks 
like an instanton.
We can choose the static gauge where the Calabi-Yau 
coordinates $X^{U}$ are identified with the world-volume coordinates $\sigma^i$.
The action $S_5$ in this gauge simplifies and becomes
\begin{equation}
S_5= T \int d^6 \sigma (\sqrt{det g_{ij}}+iC_{123456})+T\int d^6 \sigma\sqrt{det g_{ij}}
V^i \Psi_i,
\label{11}
\end{equation}
where $g_{ij}$ is now the Calabi-Yau metric and the vertex operator $V^i$ is given by 
\begin{equation}
V^i=g^{ij}\bar \Theta \Gamma_j- \frac{\epsilon^{i_1 \dots i_5 i}}{6!\sqrt{g_{ij}}}
{\bar \Theta} \Gamma_{i_1\dots i_5}.
\label{12}
\end{equation}
Now let us discuss the zero modes. This instantons
has five bosonic zero modes, four along four-dimensional 
Minkowski space and one additional zero mode along the interval.
In the low-energy limit, the fields $g_{ij}$ and $\Psi_i$ do not depend
on the interval and this additional zero mode can be integrated out to produce 
an irrelevant constant. 
The integral over the remaining bosonic zero modes is just the integral over 
four-dimensional Minkowski space. Now let us move on to the fermionic zero modes. 
We start by recalling that $\Theta$ is a Majorana spinor in eleven dimensions. 
Thus, it has thirty two real components. The Calabi-Yau background breaks 
a quarter of supersymmetry. A five-brane reduces the number of surviving supersymmetries 
by a factor of two. Thus, we find that $\Theta$ has four zero modes
in the background of the five-brane instanton. 
The equation that these zero modes is the equation 
for the supersymmetries preserved by a five-brane.
It reads
\begin{equation}
\frac{1}{6!} \Gamma^{i_1i_2i_3i_4i_5i_6}
\epsilon_{i_1i_2i_3i_4i_5i_6}\Theta =\Theta.
\label{13}
\end{equation}
Equivalently, this equation can be understood as 
the kappa supersymmetry fixing condition in the world-volume action~\eqref{6}.
Equation~\eqref{13} is just the condition that $\Theta$ has to be a chiral
spinor on Calabi-Yau manifold. This means that $\Theta$ can be written as
\begin{equation}
\Theta=\theta_{\alpha}\otimes\xi_+ \oplus {\bar \theta}_{\dot \alpha}\otimes\xi_+, 
\label{14}
\end{equation}
where $\xi_+$ is the covariantly constant spinor of positive chirality
on Calabi-Yau manifold. 
The spinors $\theta_{\alpha}$ and ${\bar \theta}_{\dot \alpha}$
are the four-dimensional spinors representing the fermionic zero modes of the 
five-brane instanton. Of course, one can come to the same conclusion analyzing the 
fermionic equations of motion derived from the action~\eqref{6}.
The equation of motion for $\Theta$ is the Dirac equation~\cite{Kallosh}
\begin{equation}
\Gamma^i(\partial_i+\omega_i^{AB}\Gamma_{AB})\Theta=0, 
\label{15}
\end{equation}
where $\omega_i^{AB}$ is the spin connection. Since a Calabi-Yau
background breaks a quarter of suppersymetry, a solution to~\eqref{15}
has eight independent components. This solution involves fermions 
of both chirality. A supersymmetric five-brane instanton corresponds 
to a solution of positive chirality (similarly an anti-five-brane instanton
corresponds to a solution of negative chirality) and we obtain a solution 
for $\Theta$ given by equation~\eqref{14}.
The five-brane partition function can now be written as
\begin{equation}
\int d^{4}x d^2 \theta d^2 {\bar \theta} e^{-S_5}, 
\label{16}
\end{equation}
where $S_5$ is given by~\eqref{11} and $\Theta$ is given by eq.~\eqref{14}.
The existence of four fermionic zero modes makes a five-brane instanton 
different from string or open membrane instantons studied 
in~\cite{Wittend, Lima1, Lima2}. String and open membrane 
instantons are boundary instantons. They have only two 
fermionic zero modes because the boundary in heterotic M-theory reduces the number of preserved 
supercharges by two. 

The next step is to evaluate the action $S_5$. Let us start with the first term
which is purely bosonic. Upon the dimensional reduction, the Calabi-Yau metric 
$g_{ij}$ is written as follows~\cite{Lukas4}
\begin{equation}
g_{ij}=V^{1/3}\Omega_{ij}, 
\label{17}
\end{equation}
where $\Omega_{ij}$ is the reference metric and $V$ is the volume modulus. 
The field $C_{123456}$ is related to the axion. This can be shown as follows. 
Take the three-form potential $C_{MNP}$ and consider the components
$C_{\mu \nu 11}$. To obtain the axion $\sigma$, we dualize the field 
strength $G_{\mu \nu\lambda 11}$ 
\begin{equation}
G_{\mu \nu\lambda 11}=\epsilon_{\mu \nu\lambda \rho}\partial^{\rho}\sigma.
\label{18}
\end{equation}
On the other hand, by definition of $C_{123456}$, we have 
\begin{equation}
G_{\mu \nu\lambda 11}=\frac{1}{6!}
\epsilon_{\mu \nu \lambda \rho i_1 \dots i_6}\partial^{\rho}C^{i_1\dots i_6}=
\epsilon_{\mu \nu\lambda \rho}\partial^{\rho}C_{123456}.
\label{19}
\end{equation}
Comparing~\eqref{18} and~\eqref{19}, we obtain 
\begin{equation}
C_{123456}=\sigma. 
\label{20}
\end{equation}
Thus, the first term in~\eqref{11} gives
\begin{equation}
TvS, 
\label{21}
\end{equation}
where 
\begin{equation}
S=V+i\sigma
\label{21.1}
\end{equation}
and $v$ is the Calabi-Yau reference volume. 
Now let us consider the term $V^i \Psi_i$ in  eq~\eqref{11}. We split the index $i$ 
into holomorphic and antiholomorphic indices $u, {\bar u}$. Then the dimensional reduction 
of the gravitino $\Psi_u$ is
\begin{equation}
\Psi_{u}=\psi_{\alpha}\otimes \Gamma_u \xi_{+}\oplus{\bar \psi}_{\dot \alpha} 
\otimes \Gamma_u\xi_+. 
\label{22}
\end{equation}
Substituting~\eqref{14} and~\eqref{22} in~\eqref{11}
and using the following relations for the covariantly constant spinors
\begin{equation}
\Gamma_{{\bar u}}\xi_+=0, \quad \xi_+^{\dagger}\xi_+=1,
\label{23}
\end{equation}
we
find that the five-brane partition function is 
\begin{equation}
\int d^4 x 
d^2 \theta d^2 {\bar \theta} e^{-TvS}
e^{-\theta^{\alpha}\psi_{\alpha} -
{\bar \theta}_{\dot \alpha}{\bar \psi}^{\dot \alpha}}.
\label{24}
\end{equation}
To perform the integral over the fermionic zero modes, we
expand the second factor in~\eqref{24} to the fourth order. This gives
the following expression for the partition function
\begin{equation}
\int d^4 x e^{-TvS} 
(\psi^{\alpha}\psi_{\alpha})
({\bar \psi}_{\dot \alpha}{\bar \psi}^{\dot \alpha}).
\label{25}
\end{equation}
This partition function is interpreted as the four-fermionic term in the action
$\delta S$ given by eq.~\eqref{4} (or, equivalently,~\eqref{1}) with the holomorphic section
$\omega$ given by
\begin{equation}
\omega_{\bar S}^{{\ }S}=e^{-TvS}.
\label{26}
\end{equation}
Of course, this function $\omega$ has to be multiplied by 
the complex structure dependent bosonic and fermionic one-loop determinants.
Analysis of fluctuations around the zero modes
and their one-loop determinants is rather complicated. In the static gauge, the 
fluctuations involve five scalars $\delta X^{\mu}$, $\mu=0,\dots, 3$, $\delta X^{11}$,
four chiral fermions $\delta \Theta_{+}$ (the fermions of opposite 
chirality are projected out by the condition~\eqref{13}, or, equivalently,
by fixing the kappa supersymmetry) and also
the anti-self-dual
two-form, whose coupling to the world-volume we have been ignoring.
Note that the fluctuations form the tensor multiplet
propagating on the five-brane worldvolume.
Various issues about the partition function of the five-brane 
world-volume fields, including anomaly cancellations,
were studied by Witten in~\cite{Wittenfive}. 
We will not discuss them in this paper. 
Because of the one-loop determinants, there is a non-trivial
non-perturbative mixing of the moduli spaces of the volume and complex
structure multiplets. So, to be precise, eq.~\eqref{26}
defines only the $S{\bar S}$ component of $\omega$. 
 
%%%%%%%%%%%%%%%%%%%%%%%%%%%%%%%%%%%%%%%%%%%%%%%%%%%%%%%%%%%%%%%%%%%%%%%%%%%%%%%%%%%%%%%%%%%%%%%%%%%%%%%%%%%%

\section{Additional Remarks}

%%%%%%%%%%%%%%%%%%%%%%%%%%%%%%%%%%%%%%%%%%%%%%%%%%%%%%%%%%%%%%%%%%%%%%%%%%%%%%%%%%%%%%%%%%%%%%%%%%%%%%%%%%

In this section, we will discuss some additional features
of five-brane instantons. First, a generic supersymmetric heterotic M-theory 
background contains some amount of the $G$-flux. The non-vanishing componets are
$G_{(2,1,1)}, G_{(1,2,1)}$ and $G_{(2,2,0)}$. Here we are using 
notation from~\cite{Wittenstrong}. The first two indices represent 
the holomorphic and anti-holomorphic directions along the Calabi-Yau manifold
and the last index is the index along the interval. 
The fact that these components of the fluxes are consistent 
with supersymmetry means that the Dirac operator on the deformed 
compactification manifold still has solutions. To be consistent with 
Poincare symmetry in five dimensions, the variation of the eleven-dimensional 
gravitino has to have eight zero modes as in the absence of fluxes. 
This immediately implies that a five-brane wrapped on the deformed manifold 
will have four fermionic zero modes since it breaks one half of supersymmtery. 
Thus a five-brane instanton still has four zero modes and constibutes 
to the function $\omega$ as $e^{-Tv(V+i\sigma)}$ times the one-loop determinants. 
However, the flux $G_{(2,2,0)}$ is more subtle. It provides a deformation of 
the Calabi-Yau metric along the interval~\cite{Wittenstrong, Curio1, Curio2}
\begin{equation}
ds^2=e^{-f(x^{11})}\eta_{\mu \nu}dx^{\mu}dx^{\nu} +
e^{f(x^{11})}(g_{ij}dX^i dX^j +dx^{11}dx^{11}), 
\label{27}
\end{equation}
where the warp factor $e^{f(x^{11})}$ is given by 
\begin{equation}
e^{f(x^{11})}=(1+x^{11}Q)^{2/3}
\label{28}
\end{equation}
and $Q$ is the amount of the flux. For example, in the absence of five-brane wrapping
holomorphic curves, $Q$ is given by~\cite{Wittenstrong, Curio1, Curio2} 
\begin{equation}
Q=\frac{\ell_{11}^3}{32\pi^2v}\int \omega \wedge 
\left({\rm tr} F\wedge F -\frac{1}{2}{\rm tr} R\wedge R\right),
\label{29}
\end{equation}
where $l_{11}$ is the eleven-dimensional Planck length and 
$\omega$ is the Kahler form of the undeformed Calabi-Yau metric. 
Then it follows from eq.~\eqref{27} that the tension 
of the five-brane wrapping the Calabi-Yau manifold is 
$x^{11}$ dependent and, thus, such an instanton is not BPS. 
The wrapped five-brane has to move along the interval to minimize
its tension until it collides with the boundary. 
Therefore, strictly speaking, the analysis in the previous section is valid only 
if a five-brane instanton is placed in the region where the flux is zero. 
The existence of such regions was discussed in detail in~\cite{Bound}
and we will not repeat it here. 

As the second remark, let us discuss what happens 
when the five-brane is placed on top of one of the orbifold fixed planes.  
This system is different than a five-brane in the bulk. 
The boundary breaks one half of the bulk supersymmetry. 
The supercharges preserved by the boundary are given by
\begin{equation}
Q=Q_{\alpha}\otimes \xi_{+} \oplus 
{\bar Q}_{\dot \alpha}\otimes \xi_{-}.
\label{30}
\end{equation}
Comparing this with~\eqref{14} we find that the zero mode of the 
boundary five-brane are given by
\begin{equation}
\Theta=\theta_{\alpha}\otimes \xi_+.
\label{31}
\end{equation}
This means that 
a five-brane instanton will now have only two zero modes, $\theta_{\alpha}$ just like
an open membrane~\cite{Lima1, Lima2}. 
Then, one can imagine that such a configuration might contribute to 
two-fermion terms, that is to
the superpotential 
for the Calabi-Yau volume multiplet. 
This would have
have interesting effects
on moduli stabilization~\cite{CK}. 
Unfortunately, it is not clear how 
to prove (or disprove) it quantitatively.
The world-volume of a
five-brane instanton coincident with 
an orbifold fixed plane interacts with a tensionless string propagating on the Calabi-Yau 
manifold. It is not known how to describe this interaction. 

As the last remark, we will recall that 
a five-brane can dissolve in the orbifold plane through 
a small instanton transition~\cite{Wittensmall, Park, Donagi}.
The resulting configuration is a gauge instanton.
If the five-brane is wrapping the entire Calabi-Yau manifold, the dissolved
instanton configuration is an instanton on $R^{4}$. Thus, the effects
of this dissolved five-brane are the standard gauge instanton effects in field theory. 
 
%%%%%%%%%%%%%%%%%%%%%%%%%%%%%%%%%%%%%%%%%%%%%%%%%%%%%%%%%%%%%%%%%%%%%%%%%%%%%%%%%%%%%%%%%%%%%%%%%%%%%%%%%%%%%%%%%%%%%%%%%%%%

\section{Acknowledgments}

%%%%%%%%%%%%%%%%%%%%%%%%%%%%%%%%%%%%%%%%%%%%%%%%%%%%%%%%%%%%%%%%%%%%%%%%%%%%%%%%%%%%%%%%%%%%%%%%%%%%%%%%%%%%%%
The author is very grateful to Amihay Hanany and Burt Ovrut for helpful
discussions.
Research at Perimeter Institute is supported in part by the Government
of Canada through NSERC and by the Province of Ontario through MEDT.

%%%%%%%%%%%%%%%%%%%%%%%%%%%%%%%%%%%%%%%%%%%%%%%%%%%%%%%%%%%%%%%%%%%%%%%%%%%%%%%%%%%%%%%%%%%%%%%%%%%%%%%%%%%%%%%%%%%%%

%%%%%%%%%%%%%%%%%%%%%%%%%%%%%%%%%%%%%%%%%%%%%%%%%%%%%%%%%%%%%%%%%%%%%%%%%%%%%%%%%%%%%%%%%%%%%%%%%%%%%%%%%%%%%%

\end{document}